\documentclass[12pt]{article}
  
\usepackage{epsfig}
\usepackage{calrsfs,bbm,bm}
\usepackage{color,cancel,graphicx,mathrsfs}
\usepackage{amsmath}
\usepackage{amssymb}
 
\usepackage{exscale}

\usepackage{hyperref}

\usepackage{nicefrac}

\def\a0size{6}

\newcommand{\lsi}{\raise0.3ex\hbox{$<$\kern-0.75em\raise-1.1ex\hbox{$\sim$}}}

\newcommand{\gsi}{\raise0.3ex\hbox{$>$\kern-0.75em\raise-1.1ex\hbox{$\sim$}}}

\newcommand{\lsim}{\mathop{\lsi}}
\newcommand{\gsim}{\mathop{\gsi}}
\renewcommand{\vec}[1]{{\bf #1}}

\newcommand{\be}{\begin{equation}}
\newcommand{\ee}{\end{equation}}
\newcommand{\baed}{\begin{aligned}}
\newcommand{\eaed}{\end{aligned}}

\begin{document} 

\setlength{\baselineskip}{0.6cm}
\newcommand{\figysize}{16.0cm}
\newcommand{\figtopspace}{\vspace*{-1.5cm}}
\newcommand{\figbottomspace}{\vspace*{-5.0cm}}
  

\begin{flushright}
BI-TP 2013/18
\\
November 2013
\\
\end{flushright}
\begin{centering}
\vfill

{
\centerline{ \Large \bfseries
     Non-relativistic leptogenesis
    }
}

\vspace{1cm}

Dietrich B\"odeker\footnote{bodeker@physik.uni-bielefeld.de}
and
Mirco W\"ormann\footnote{mwoermann@physik.uni-bielefeld.de}

\vspace{.6cm} { \em 
Fakult\"at f\"ur Physik, Universit\"at Bielefeld, D-33615 Bielefeld, Germany
}

\vspace{2cm}
 
{\bf Abstract} 

\end{centering}
 
\vspace{0.5cm}
\noindent
In many phenomenologically interesting models of thermal leptogenesis
the heavy neutrinos are non-relativistic when they decay and
produce the baryon asymmetry of the Universe. We
propose a non-relativistic approximation for the corresponding rate
equations in the non-resonant case, and a systematic way for computing relativistic
corrections.  We determine the leading order coefficients in these
equations, and the first relativistic corrections.  The
non-relativistic approximation works remarkably well. It appears to be
consistent with results obtained using a Boltzmann equation taking into account the
momentum distribution of the heavy neutrinos, while being much
simpler. We also compute radiative corrections to some of the coefficients 
in the rate equations. Their effect is of order 1\% in the regime 
favored by neutrino oscillation data.
We obtain the correct leading order lepton number washout rate in this regime, 
which leads to large ($\sim 20$\%) effects compared to previous
computations.

\vspace{0.5cm}\noindent

 
\vspace{0.3cm}\noindent
 
\vfill \vfill
\noindent
 

 
\section{Introduction and motivation}
\label{sec:intro} 

Leptogenesis \cite{fukugita} (for reviews, see e.g. 
  Refs.~\cite{buchmuller-origin,davidson-review})
is an appealing scenario which can
explain the baryon asymmetry of the Universe. It requires 
the existence of sufficiently heavy
right-handed, or sterile neutrinos $ N _ i $ in addition
to the particles of the Standard Model. Their Yukawa couplings
with the Higgs and the Standard Model leptons have to contain $ CP $-violating
phases. When their interactions occur out of thermal 
equilibrium, they can produce a lepton asymmetry $ L $, which, due to
$ B + L $ violating sphaleron processes, also leads to a
baryon asymmetry $ B $. 

During leptogenesis most interactions are either much faster or 
much slower than the expansion 
of the Universe. Therefore most
physical quantities are either very close to their corresponding 
thermal equilibrium values, or constant. 
A few  quantities  evolve roughly at the same rate as the Hubble rate, and 
can therefore deviate significantly from their equilibrium values. 
Among them are the number density of the heavy neutrinos and 
the $ B - L $ asymmetry, because they are only changed by the heavy neutrino 
Yukawa interaction. 

However,  not only  the number density of heavy neutrinos,
but also their momentum  distribution is out of equilibrium, because 
the kinetic equilibration is also due to the Yukawa interactions.
Nevertheless, in the standard approach \cite{luty}  kinetic equilibrium is assumed.
In addition, Maxwell-Boltzmann statistics is used. 
In order to account for the correct momentum
distribution Boltzmann equations for the
distribution function have been used in Refs.~\cite{basboll,garayoa,hahn}, 
together with quantum statistics, 
and significant differences compared to the standard approach were reported. 

In the standard approach both $ 1 \leftrightarrow  2 $ decays
and inverse decays of the right-handed neutrinos, 
as well as $ 2 \leftrightarrow  2 $ scattering
processes are taken into account, and both were found to be important.
Recently the production rate of right-handed neutrinos 
of mass $ M _ N $, 
which is one ingredient of the rate equations describing leptogenesis,
were studied at next-to-leading order in the Standard Model 
couplings~\cite{salvio,laine-nlo,laine-nlo-relativistic}.
It was pointed out that the $  2 \leftrightarrow  2 $ 
processes are only part of
the next-to-leading order contribution and that $ 1 \leftrightarrow 3
$ scattering and virtual corrections to the (inverse) decays
contribute at the same order. 
If they are taken into account the large infrared contributions cancel 
\cite{salvio}. Therefore a complete next-to-leading
order calculation would be important. 
 Results for the production rate were obtained in the 
limit $ T \ll M _ N $, together with $ T/M _ N $ corrections 
in \cite{salvio,laine-nlo}, and
more recently also in the regime $ T \sim M _ N $  \cite{laine-nlo-relativistic}. 

A measure for the interaction rate of right-handed neutrinos relative
to the Hubble rate $ H $ at the time when they become non-relativistic is  
the washout factor
\begin{align} 
   K \equiv \left . \frac{ \Gamma  _ 0} { H } \right | _ { T = M _ N } 
   \label{K}
   , 
\end{align} 
where $ \Gamma _ 0 $ is the total tree level
decay rate. In the so-called 
strong washout regime $ K \gg 1 $  the right-handed neutrinos are close 
to equilibrium  around $ T \sim M _ N $,  any pre-existing asymmetry is washed out, 
and the final asymmetry is mainly created when 
$ T $ has dropped below $ M _ N $.
At this point the right-handed
neutrinos are non-relativistic, since they are close to 
equilibrium.
It turns out that the strong washout regime corresponds 
to a mass range of the light neutrinos
which is favored by atmospheric and solar neutrino oscillation data
\cite{buchmuller-origin}. 
Therefore the non-relativistic regime is of particular interest for
leptogenesis. A first principle treatment of leptogenesis, which 
could allow for a complete next-to-leading order calculation can
be expected to be simpler in this regime than in the fully 
relativistic case.

In this paper we obtain the rate equations for leptogenesis in the non-relativistic limit,
and we propose a systematic expansion to compute relativistic corrections.   
This way one can  systematically control the accuracy of the non-relativistic
approximation, and one can obtain information about the momentum
distribution and thus about the deviation from kinetic equilibrium.
The rate equations are valid to all orders in the Standard Model couplings, 
and a  complete  next-to-leading order calculation in these couplings 
should be feasible. We take a first step in this direction by computing
the corrections for some coefficients in the rate equations, thereby consistently
including temperature dependent gauge and top Yukawa interaction in a
leptogenesis calculation for the first time. 
We restrict ourselves to the case of hierarchical Majorana masses for the
right-handed neutrinos, and we only consider the leading order in the
right-handed neutrino Yukawa couplings. 

This paper is organized as follows. 
In Sect.~\ref{sec:non-relativistic} we 
obtain the equations for leptogenesis  in the non-relativistic limit
and to all orders in the Standard Model couplings. In Sect.~\ref{sec:lo} we compute
the coefficients in these equations at leading order in the Standard Model couplings.
The conditions
for thermal (non-)equilibrium are discussed in Sect.~\ref{sec:neq}.
The first relativistic corrections to the equations of Sect.~\ref{sec:non-relativistic} 
are obtained in Sect.~\ref{sec:relativistic}.  In Sect.~\ref{sec:radiative} 
we use the results of \cite{laine-nlo} to compute some of the coefficients 
at next-to-leading order in the Standard Model couplings. 
Numerical results are shown in Sect.~\ref{sec:numerical}, and 
the conclusions are in Sect.~\ref{sec:discussion}.

\section{Leptogenesis in the non-relativistic limit}
\label{sec:non-relativistic}

We assume that the mass of the right-handed neutrinos is hierarchical and that
we only need to take into account the lightest one which we denote by $ N $.
We work at leading order in the Yukawa couplings of the right-handed neutrinos.
We also assume that the interactions of the right-handed neutrinos are  
much less frequent than gauge interactions. \footnote{\label{stopping}The rate for
large angle scattering via gauge interactions is of order 
$  g ^ 4 T $ modulo logarithms, where $ g $ denotes a gauge
coupling. When a non-relativistic neutrino  with mass $ M _ N $ decays, 
its decay products originally have energy of approximately   
$ M _ N / 2 \gg T $. They lose 
their energy within a stopping time of order $ t _ {\rm stop } $ 
with $ t _ {\rm stop } ^{ -1 } 
\sim \sqrt{ T/M _ N  }\, g ^ 4 T  $ 
modulo logarithms \cite{baier,arnold-stopping}.}
Then on time scales which are relevant
to leptogenesis the Standard Model particles are in kinetic 
equilibrium. Finally we assume that the Standard Model Yukawa interactions are either much
faster than the right-handed neutrino reaction rates so that they are in 
thermal equilibrium, or that they are much slower. \footnote{If they 
occur at roughly the same rate they both have to be included in the
equations for leptogenesis \cite{beneke-flavored}.} 

We assume that only one lepton flavor is involved. This
is the case, for instance, at high temperature, when all Yukawa interactions
of the Standard Model leptons can be neglected. At lower
temperature the lepton Yukawa interactions become important, 
and in general one has to deal with several asymmetries \cite{barbieri}.
However, for certain values of the Yukawa coupling matrix
one may still be approximately in a one-flavor regime. 

We consider non-relativistic right-handed neutrinos with a characteristic
velocity $ v \ll 1 $. In a first approximation their motion can be
neglected.  Then the out-of-equilibrium state is fully specified by the
number densities $ n _ N $ and $ n _ { B - L } $. In a non-expanding
Universe the time derivative of these densities would only depend
on their deviations from their equilibrium values and on the temperature. 
These considerations imply that the time
evolution of this system in an expanding Universe is governed by the equations
\begin{align}
   \left ( \frac{ d   } { d t }  + 3 H \right )  n _ N 
       \, \, \, & = {} - 
    \Gamma _ N  \, \, ( n _ N - n _ N ^ {\rm eq}  ) 
       +   \Gamma  _ { N, B-L } \, n _ { B - L } 
   \label{dnNdt}
   ,
\\
   \left ( \frac{ d  } { d t }   + 3 H  \right )   n _ { B - L }  
      & = 
   \, \, \Gamma  _ {B-L, N } 
   ( n _ N - n _ N ^ {\rm eq}  ) 
    - 
    \Gamma  _ {B - L } 
    \, n _ { B - L }
   \label{dnLdt}
   .
\end{align} 
The  coefficients $ \Gamma  _ i $ 
only depend on the temperature of the ambient plasma.
$ \Gamma  _ N $ describes how quickly the number density of heavy neutrinos
approaches its equilibrium value, and 
$  \Gamma  _ {B - L }  $ describes
the dissipation, or washout, of $ B - L $. 
$ \Gamma  _ {B-L, N }  $ 
parametrizes how the deviation of $ n _ N $ from equilibrium
creates an asymmetry of $ B - L $. This reaction, as well as the one
described by 
$ \Gamma  _ {N, B-L }  $ violate $ CP $.
We stress that these
equations are valid to all orders in the Standard Model couplings.
The only corrections are due to
relativistic effects of the motion of $ N $. \footnote{For large deviations
from equilibrium there could also be non-linear terms. These are
expected to be higher order in the $ N $-Yukawa couplings.}

For leptogenesis one is usually interested in an initial state
with vanishing $ n _ { B - L } $, which is then generated by 
non-vanishing $ n _ N - n _ N ^{ \rm eq } $ as described
by Eq.~(\ref{dnLdt}). The coefficient $ \Gamma  _ { B-L, N } $ 
contains a small parameter $ \epsilon  $ (cf. Eq.~(\ref{epsilon}))
resulting in $ n _ { B-L } \ll n _ N - n _ N ^ {\rm eq }  $. 
We expect the coefficient $ \Gamma  _ { N, B-L } $ to contain
a similarly small factor, so that the second term 
on the right-hand-side of Eq.~(\ref{dnNdt}) can be safely neglected.

\section{Leading order coefficients}
\label{sec:lo}

We would like to determine the coefficients in Eqs.~(\ref{dnNdt}), 
(\ref{dnLdt})
at leading order in the Standard Model couplings. We expect 
that at this order
they can be obtained from a Boltzmann
equation containing the $1 \leftrightarrow 2 $ scattering processes.

We work at leading order in the Yukawa coupling
of right handed neutrinos, and  we therefore neglect processes
which change lepton number by two units via the 
exchange between 
Standard Model particles. The corresponding rates are suppressed
by an additional factor $ h ^ 2 $, where $ h $ is a generic
Yukawa coupling of the right handed neutrinos (see below).

Since leptons and Higgs bosons are in kinetic equilibrium,
their phase space distributions
can be parametrized by the temperature and by
chemical potentials. In the Boltzmann-equation 
for $ f _ N ( t, p )$ with $ p
\equiv | \vec p | $ we can neglect the chemical potentials of Higgs and
leptons because they are of the order of the $ B - L $ asymmetry (recall that
$ n _ { B-L } \ll n _ N - n _ N ^ {\rm eq }  $). We use
Boltzmann statistics instead of Fermi- or Bose-statistics for leptons
and Higgs, because they have energies of order $ M _ N / 2 \gg T$ when they
produce an $ N $. Furthermore, we neglect the Pauli blocking
and Bose enhancement factors. The corresponding errors are suppressed 
with $ \exp [ - M_N/ T   ]  $.  After performing the phase space integral 
the Boltzmann equation becomes
\begin{align} 
  \left (  \partial _ t - H p \partial _ p \right )  f _ N  = 
    \frac{ M _ N \Gamma _ 0} { E _ N } \left ( e ^{ - E _ N / T } - f _ N \right  ) 
    \label{boltzmannN}
    ,
\end{align} 
where $ \Gamma  _ 0 $ is the total tree-level decay rate of the 
heavy neutrinos, and $ E
_ N = ( \vec p ^ 2 + M _ N ^ 2 ) ^{ -1/2 } $ is their energy. 
We consider the Yukawa  interaction
\begin{align}  { \cal L } _ { N { \rm Yuk  } } 
= h_{ij} \overline{N _i 
                  }{\widetilde \varphi}^\dagger 
   \ell_{
   j}
   + \mbox{ h.c. } 
\end{align} 
with the left-handed Standard Model lepton doublets $ \ell _ j $ and the isospin
conjugate $ {\widetilde \varphi} \equiv i \sigma ^ 2 \varphi
^ \ast  $ of the Higgs doublet $ \varphi  $. Then
the decay rate of the lightest right-handed neutrino $ N \equiv N _ 1 $ is  
\begin{align} 
 \Gamma  _ 0 = \frac{ | h _ { 11 } | ^ 2 M _ N } { 8 \pi  } 
 \label{Gamma0}
 ,
\end{align} 
if $ \ell _ 1 $ is the lepton produced in the $ N $-decays (see Sect.~\ref{sec:non-relativistic}).

One can obtain an equation for $ n _ N = (2 s _ N + 1 ) ( 2 \pi ) ^{ -3 }
\int d ^ 3 p f _ N $, where $ s _ N = 1/2 $ is the spin of $ N $, 
by integrating Eq.~(\ref{boltzmannN}) over $ \vec p $. However, 
the loss term on the right-hand side would not only depend on the number density
$ n _ N $, but also on the momentum spectrum of the right-handed neutrinos.
In most papers on leptogenesis this problem is circumvented
by assuming that the $ N $'s are in kinetic equilibrium. Then the
deviation from thermal equilibrium is completely specified by $ n _ N
$.  Here we do not assume kinetic equilibrium. Instead, we make use of the
fact that the $ N $'s are non-relativistic,
and approximate  the factor $ 1/E _ N $ in front of
the bracket by $ 1/M _ N $. For the first coefficient in Eq.~(\ref{dnNdt})  
we thus obtain
\begin{align} 
   \Gamma  _ N = \Gamma  _ 0 
   \label{GammaN}
\end{align} 
at leading order in the Standard Model couplings.
Corrections to Eq.~(\ref{GammaN}) will be discussed
in Sect.~\ref{sec:radiative}. 

Now we determine the coefficients in Eq.~(\ref{dnLdt}).
In many cases \footnote{Eq.~(\ref{boltzmannN}) is an exception.}
the Boltzmann equation for a given particle species contains 
collision terms which are much larger than the collision terms 
contributing to leptogenesis
itself. The processes which give these large contributions
are called 
spectator processes 
\cite{buchmuller-spectator}. Their role is to bring the 
quantum numbers which are not conserved into thermal 
equilibrium.
Important examples are scattering by gauge boson exchange, 
the anomalous quark-chirality changing QCD-sphalerons
and the  anomalous $ B + L $ changing electroweak sphaleron processes.
Which spectators are in equilibrium depends on the temperature 
(cf.\ Table~\ref{tab:spectators}).  
All spectator processes conserve $ B - L $. Thus, after 
adding the Boltzmann equations for all quarks 
and antiquarks, weighted by their respective baryon number, 
subtracting the equations for leptons and antileptons, weighted by
their lepton number, and then integrating over momentum the
spectator collision terms drop out. The only violations of 
$ B - L $ are due to the interactions of the right-handed neutrinos. 

The decays of the right-handed neutrinos contribute to the
coefficient 
$ \Gamma _ { B - L, N } $.  The asymmetry arises because the decay
rate of $ N $ into matter differs from the one into antimatter, which
is quantified by a non-zero value of
\begin{align} 
   \epsilon  \equiv 
   \frac{
   \Gamma  ( N \to \varphi  \ell ) 
     - \Gamma  ( N \to \overline{ \varphi  } \overline{ \ell } )
     }
     {
   \Gamma  ( N \to \varphi  \ell ) 
     + \Gamma  ( N \to \overline{ \varphi  } \overline{ \ell } )
     }
     \label{epsilon}
   .
\end{align} 
Therefore 
\begin{align} 
  \Gamma  _ { B - L, N } 
  = { \epsilon  \Gamma  _ 0 }
   \label{GammaB-LN}
\end{align}
at leading order in the Standard Model couplings.

Inverse decays of the heavy neutrinos contribute to the washout term 
in Eq.~(\ref{dnLdt}) at leading order. At higher orders in the
$ h _ { ij } $ 
there are also $ 2 \leftrightarrow  2 $
scattering processes with a right handed neutrino exchange
which change lepton number by two units. Unlike 
the inverse decays these are not exponentially suppressed 
in the non-relativistic regime (see below) which may compensate
the suppression by
the Yukawa couplings and powers of $ T / M _ N $. 
The washout factor (\ref{K}) is of order 
$ K \sim  h _ { ij } ^ 2 m _ {\rm Pl } / M _ N $, where $ m _ {\rm Pl } $ 
denotes the Planck mass. Thus at fixed $ K $,  $ h _ { ij } $ increases
with increasing $ M _ N $.  Therefore at very large $ M _ N $ 
these processes can become important, but we do not consider them here. 

Like in the computation of $ \Gamma  _ N $ 
one can use Maxwell-Boltzmann statistics and neglect
Pauli blocking and Bose enhancement terms, which gives
\begin{align} 
   \Gamma _ { B-L }\, n _ { B - L } 
   = 
   \int \prod _ { a = N, \ell, \varphi  } \frac{ d ^ 3 p _ a } 
   { 2 E _ a ( 2 \pi  ) ^ 3 } 
   ( 2 \pi  ) ^ 4 \delta  ( p _ \ell + p _ \varphi  - p _ N ) 
   \left ( f _ \ell f _ \varphi  - f _ { \overline{\ell } } 
      f _ { \overline{ \varphi  } } \right )
    \sum | { \cal M } _ 0| ^ 2 
    \label{coll}
    .
\end{align}  
Here $ \sum | { \cal M } _ 0 | ^ 2 = 16 \pi  M _ N \Gamma  _ 0 $ 
(see Eq.~(\ref{Gamma0})) is the
tree level matrix element squared for the process $ \ell \varphi  \to N $ summed
over all spins of $ N $ and all isospin components of $ \ell $.
Since we consider small deviations from equilibrium we can expand 
in the chemical potentials $ \mu  _ \ell $ and $ \mu  _ \varphi  $ 
and keep only the linear order, 
\begin{align}  f _ \ell f _ \varphi  - f _ { \overline{\ell } } 
      f _ { \overline{ \varphi  } } \simeq 2 e ^{ - E _ N / T } 
   ( \mu  _ \ell + \mu  _ \varphi  ) /T 
   \label{ff}
   .
\end{align} 
The chemical potentials 
\footnote{Oftentimes the chemical potential of the Higgs bosons is ignored.} 
are proportional to $ n _ { B-L }$.
Their precise relation is temperature dependent, because it depends on
which spectator processes are in effect. 
One can introduce coefficients $ c _ \ell
$ and $ c _ \varphi   $ to parametrize the relation of lepton and Higgs asymmetries 
to $ B - L $, 
\footnote{We thank Sacha Davidson
  for clarifying remarks regarding these relations.}
\begin{align} 
   n _ \ell - n _ { \overline{ \ell } } =& - c _ \ell  \, n _ { B - L } 
   ,\label{cell} 
   \\
   n _ \varphi  - n _ { \overline{ \varphi  } } = & - c _ \varphi  \, n _ { B - L } 
   \label{cphi}
   .
\end{align} 
Results for $  c _ \ell $ and $ c _ \varphi   $ for various temperature
regimes can be found in Table 1 of Ref.~\cite{davidson-review}.
To evaluate the left-hand sides of Eqs.~(\ref{cell})
and (\ref{cphi}) one can expand the  Fermi- and Bose-
distributions for leptons and Higgses up to linear order in the chemical
potentials, and then integrate over momenta. 
Note that the momentum integrals are saturated at $ p \sim
T $. That means that, unlike above, one cannot use the Maxwell-Boltzmann
distribution. Using the Fermi- and Bose-distributions instead one 
finds 
\begin{align} 
      \mu  _ \ell  & = \frac{ 3 c _ \ell   } 
   { T ^ 2 } \, n _ { B - L }
   \label{muell}
   , \\
   \mu  _ \varphi  & = \frac{ 3 c _ \varphi  } 
   { 2 T ^ 2 }\, n _ { B - L }
   \label{muphi}
   .
\end{align}
Combining Eqs.~(\ref{coll}), (\ref{ff}), (\ref{muell}), and (\ref{muphi}) 
one finally obtains for the washout rate
\begin{align} 
  \Gamma  _ { B - L } 
  = \frac{ 3 } { \pi  ^ 2 } 
  \left ( c _ \ell  + \frac{ c _ \varphi  } { 2 } \right ) 
  z ^ 2 K _ 1 ( z ) \Gamma  _ 0 
  \label{GammaB-L}
   .
\end{align} 
Here $ K _ 1 $ is the modified Bessel function of the second kind and 
\begin{align}
   z \equiv \frac{ M _ N } { T }
   \label{z}
   .
\end{align} 
Note that  we did not make a 
non-relativistic approximation for  $ \Gamma _ {B - L } $. Our result
differs from 
Ref.~\cite{buchmuller-pedestrians} (where $ c _ \ell = 1 $, $ c _ \varphi  =0 $ 
was used) and \cite{davidson-review} by a factor $ 12/\pi  ^ 2 $. This 
is because we have used quantum statistics to obtain Eqs.~(\ref{muell}) and
(\ref{muphi}). It is suggestive to use 
classical statistics for Eqs.~(\ref{coll}) and (\ref{ff}), because
the lepton number washout is due to leptons and Higgs bosons
with energy of order $ M _ N /2 \gg T $. However, the kinetic 
equilibration rate (see footnote \ref{stopping} on page \pageref{stopping})
is much larger than the interaction rate of the heavy neutrinos. Therefore
the relation between 
the asymmetry and the chemical potentials is determined by the 
bulk of kinetically equilibrated leptons and Higgs bosons, which have momenta of order $ T $. 

Let us finally comment on the washout due to $ \Delta L = 2 $ processes 
mediated by the exchange of a right-handed Majorana neutrino. Unlike 
Eq.~(\ref{GammaB-L}) the corresponding rate is not Boltzmann
suppressed with $ e ^{ -M _ N /T } $ for $ T \ll M_N $. On the other hand, 
the $ \Delta L = 2 $ rate is of order $ h ^ 4 $. For a given value
of $ K $ large $ h ^2 $ means large $M _N $. Thus the $ \Delta L = 2 $
processes can be important for very heavy right-handed neutrinos. 
They were found to be neglible for 
$ M _ N \lesssim 10 ^ {13} $ GeV \cite{davidson-review}.

\begin{figure}[t]
  \centerline{
    \epsfxsize=10cm
    \epsffile{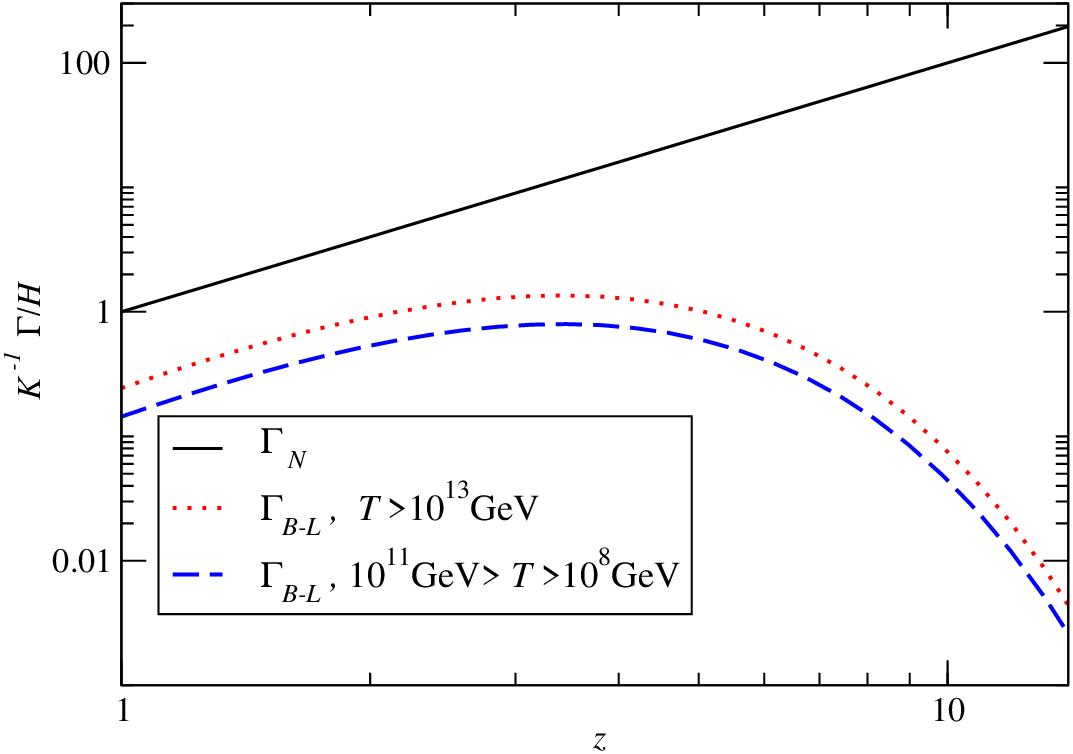}
  } \vspace{0cm}
  \caption[a]{
    \label{fg:ratio}
        Ratio of reaction rates in Eqs.~(\ref{dnNdt}), (\ref{dnLdt}) 
        and the Hubble rate, normalized to the washout factor $ K $. 
        $ \Gamma  _ N/H $ 
        (full black line) increases with $ z $, implying that abundance of 
        right-handed neutrinos gets closer and closer to equilibrium with
        increasing time, or, since the equilibrium density drops exponentially, 
        gets closer and closer to zero. On the other hand, $ \Gamma  _ { B-L }/H $
        (red dotted and blue dashed line)
        first increases, and then drops exponentially, implying that 
        $ B - L $ freezes out. 
          }
\end{figure}

\section{Interaction rates and thermal (non-)equilibrium}
\label{sec:neq}

The right-handed neutrinos are out of equilibrium if their interaction
rate is smaller than the Hubble rate. When $ T \lsim  M _ N $, 
the interaction rate can be estimated as $ \Gamma  _ 0 $. Therefore
the ratio of the two rates is 
\begin{align}  
   \frac{   \Gamma  _ N  } {  H } \simeq 
   \frac{ \Gamma  _ 0 }{ (T ^ 2/ M _ N ^ 2 ) H | _ { T=M _ N }   } 
   =   \left ( \frac{ M _ N } { T  } \right ) ^ 2 K
   \label{Nratio} 
   , 
\end{align} 
i.e., the deviation from equilibrium becomes smaller
when the temperature decreases. It also means that the deviation from 
kinetic equilibrium becomes smaller.
On the other hand, the washout rate is of the same order as $ \Gamma  _ N $ 
 for $ T \sim M _ N $,
while  for $ T \ll M _ N $
\begin{align}
   \frac{ \Gamma  _ { B - L } } { H } \sim
   \left ( \frac{ M _ N } { T } \right ) ^{ 7/2}
   e ^{ - M _ N / T }  K
   \label{Lratio}
   . 
\end{align} 
Thus  $ \Gamma  _ { B - L } / H $ should first increase
when the temperature falls below $ M _ N $, and should then drop sharply due to the exponential.
Fig.~\ref{fg:ratio} shows the ratios $ \Gamma  _ N /H $ and $ \Gamma  _ { B-L }/H $, 
normalized
to the washout factor $ K $. The ratio $ \Gamma  _ { B-L } /H $ indeed
has a maximum between $ z = 3 $ and $ z = 4 $. Thus one should expect that
this is the most important $ z $-range for leptogenesis. Futhermore, for $ K \gsim  1 $
the deviations from chemical and kinetic equilibrium should be small in 
this regime. Note that for
$ z \gg 1 $ there is a hierarchy between the scale for $ B - L $ dissipation
and of the heavy neutrino equilibration, $   \Gamma  _ { B-L } \ll \Gamma  _ N  $.

\section{Relativistic corrections}
\label{sec:relativistic}

There are corrections to Eqs.~(\ref{dnNdt}), (\ref{dnLdt}) due to the
motion of the right-handed neutrinos. In general their momentum
distribution will also deviate from thermal equilibrium. 
We expect the first correction to
be related to the kinetic energy density. There can also be relativistic 
corrections related to the spin of the right-handed neutrinos, which
we do not consider here. \footnote{We thank M.~Laine for pointing this out.}
At leading order in the gauge and Yukawa couplings we expect that 
the correct coefficients in the kinetic equations can be obtained
from Boltzmann equations which contain only the decay and inverse
decay processes. 

We have 
$ E _ N = M _ N [  1 + O ( v ^ 2 ) ]  $ where $ v $ is a typical
velocity of the $ N $. In thermal equilibrium
$ \vec p ^ 2/M _ N \sim T $ and therefore $ v ^ 2 \sim T/M  _ N
\sim 1/z $.
We expand  the factor $ 1/E _ N $ in front of
the bracket in Eq.~(\ref{boltzmannN}) up to order $ v ^ 2 $, which gives
\begin{align} 
  \left ( \frac{ d } { d t } + 3 H \right ) 
  n _ N = \Gamma _ N \left ( n _ N ^  {\rm eq } - n _ N \right ) 
  + \Gamma  _ { N,  u  } 
     \left (  u - u ^  {\rm eq } \right )
   \label{dnNdtNLO}
   .
\end{align} 
Here
\begin{align}
   u \equiv \frac{ 1 } { M _ N} (2 s _ N + 1 ) \int 
   \frac{ d ^ 3 p } { ( 2 \pi  ) ^ 3 } \frac{ \vec p ^ 2 } { 2 M _ N} 
   f _ N 
\end{align} 
is the kinetic energy density of the heavy neutrinos devided by their mass. 
At lowest order 
\begin{align}
   \Gamma  _ { N,  u } = \Gamma  _ 0
   .
\end{align} 
The equation
for $ u $ is obtained by multiplying Eq.~(\ref{boltzmannN}) with $ \vec p ^ 2 $
and integrating over $ \vec p $. At leading order in $ v $ this gives
\begin{align} 
  \left ( \frac{ d } { d t }  + 5 H \right ) u 
 =  \Gamma  _ u \left ( u ^  {\rm eq } - u \right ) 
 \label{dudt}
 , 
\end{align} 
with $ \Gamma_u = \Gamma  _ 0 $ at leading order in $ T/M _ N $ and 
in the Standard Model couplings.
The equation for the asymmetry is modified accordingly,
\begin{align} 
   \left ( \frac{ d } { d t } + 3 H \right ) n _ { B - L }
   = 
   \Gamma  _ { B - L, N } \left ( n _ N  - n _ N ^  {\rm eq } \right ) 
   + 
   \Gamma  _ { B - L, u } \left (      u - u ^  {\rm eq } \right )
   - 
   \Gamma  _ { B - L } n _ { B - L } 
   \label{dnLdtNLO}
   .
\end{align} 
Using the same expansion as for Eq.~(\ref{dnNdtNLO}) 
one finds at leading order (cf.\ Eq.~(\ref{GammaB-LN}))
\begin{align} 
    \Gamma  _ {B-L,  u } = \epsilon  \Gamma  _ 0 
   .
\end{align}

\section{Radiative corrections}
\label{sec:radiative}

At leading order only the decays and inverse decays of right-handed neutrinos contribute
to the coefficients in Eqs.~(\ref{dnNdt}) and  (\ref{dnLdt}). In many papers
$ 2 \leftrightarrow 2 $ scattering processes which involve Standard Model 
interactions have been included as well, and
their effects were found to be substantial. Recently \cite{salvio} it
was pointed out, that such calculations are not complete because they only
contain $ 2 \leftrightarrow 2 $ particle scattering, while ignoring $
1 \leftrightarrow 3 $ processes and virtual corrections to the leading order $ 1
\leftrightarrow 2 $ scattering.  Infrared contributions
which were found earlier were shown to cancel in the complete answer.

Here we show how  radiative corrections can be incorporated into our approach. 
The equilibrium densities $ n _ N ^ { \rm eq } $ and $ u ^ { \rm eq }
$ do not receive radiative corrections from Standard Model interactions.
Therefore they only affect the coefficients $ \Gamma
_ i $ in Eqs.~(\ref{dnNdtNLO}), (\ref{dudt}), and (\ref{dnLdtNLO}).

Radiative corrections to the production rate $ d n _ N / d t $ at
vanishing $ n _ N $ have been computed 
in Refs.~\cite{salvio,laine-nlo}.  In \cite{laine-nlo} also the 
radiative corrections to the differential rate  $ \partial f _ N /
\partial t $ at $ n _ N = 0 $ have been computed. It can be written 
as  
\begin{align} 
   \left . 
     \frac{ \partial  f _ N ( t, \vec p ) } { \partial t } 
     \right | _ { f _ N = 0 } 
   = 
   f _ {\rm F } ( E _ N ) \Gamma  _ 0 \frac{ M _ N } { E _ N } 
   \left \{ a + \frac{ \vec p^ 2 } { M _ N ^ 2 } \,b 
     + O \left ( \frac{ \vec  p ^ 4 } { M _ N ^ 4 } \right ) 
    \right \} 
   \label{radcorr} 
   .
\end{align} 
The coefficients $  a $, $ b $ are temperature-dependent and can be expanded
in $ T / M _ N $.

Assuming that Eq.~(\ref{dnNdt}) is still valid for $ n _ N = 0 $,
one can read off the  radiative corrections to
the coefficients in the kinetic equations for $ n _ N $ (Eq.~(\ref{dnNdtNLO})),
and for $ u $ (Eq.~(\ref{dudt})) from Eq.~(\ref{radcorr}). 
Expanding the factor $ 1/E _ N $ as in Eq.~(\ref{boltzmannN}), and 
integrating over $ \vec p $ we obtain
\begin{align} 
  \Gamma  _ N = \Gamma _ u = a \Gamma  _ 0   , \qquad 
  \Gamma  _ { N , u } =    ( a - 2 b ) \Gamma  _ 0
   . 
   \label{Gamma-w-radcorr}
\end{align} 
Radiative corrections that would enter the coefficients in Eq.~(\ref{dnLdtNLO}) 
for the asymmetry have not been computed so far.

The  coefficients in Eq.~(\ref{radcorr}) are~\footnote{ The $ O ( g ^ 2 ) $ and
the  $ O ( \lambda  T ^ 2 /M _ N ^ 2 ) $ 
corrections in $ a $ were first computed in Ref.~\cite{salvio}. The $ O ( g ^ 2 ) $
contributions in Refs.~\cite{salvio} and \cite{laine-nlo}  agree, but the
 $ O ( \lambda  T ^ 2 /M _ N ^ 2 ) $ in Ref.~\cite{salvio}
is 4 times larger than the one
in Ref.~\cite{laine-nlo} (see footnote \ref{fuss} on page \pageref{lambda}).
For our numerical results in Sect.~\ref{sec:numerical} we have used the expression in  
Ref.~\cite{laine-nlo}. The other terms in 
$ a $ and $ b $ were computed in Ref.~\cite{laine-nlo}.}

\begin{align} 
 a = {} & 
   1 - \frac{\lambda T^2}{M_N^2} 
   \, - \, 
   |h_t|^2 
   \biggl[      
     \frac{21}{2(4\pi)^2} + 
     \frac{7\pi^2 }{60}
      \frac{T^4}{M_N^4} 
   \biggr]
     + \, 
   (g_1^2 + 3 g_2^2)
   \biggl[
     \frac{29}{8(4\pi)^2} - 
     \frac{\pi^2 }{80}
     \frac{T^4}{M_N^4}  
     \biggr]
        \nonumber \\ & {}
     + O \left (  g ^ 2 \frac{ T ^ 6 } { M _ N ^ 6 } , 
      g ^ 3 \frac{ T ^ 2 } { M _ N ^ 2 }\right ) 
    \label{a}
    \,
    ,
 \end{align} 
 \begin{align}
   b = & 
   \, - \, \left [ 
   |h_t|^2 
     \frac{7\pi^2 }{45}
     + \, 
   (g_1^2 + 3 g_2^2)
     \frac{\pi^2 }{60}
     \right ]  \frac{T^4}{M_N^4} 
     + O \left (  g ^ 2 \frac{ T ^ 6 } { M _ N ^ 6 } , 
      g ^ 3 \frac{ T ^ 2 } { M _ N ^ 2 }\right ) 
     \label{b}
     ,
\end{align} 
where $ h _ t $ is the top Yukawa coupling, $ g _ 2 $, $ g _ 1 $ are the
weak SU(2) and U(1) gauge couplings, 
$ \lambda = m _ {\rm Higgs }  ^ 2 G _ {\rm F } /\sqrt{ 2 } $ 
is the Higgs self-coupling~\footnote{\label{fuss}This definition 
of $  \lambda $   is the same as the one used
in Ref.~\cite{laine-nlo}.  Ref.~\cite{salvio} writes the rate in terms of $ \lambda _ h \equiv 
 m ^ 2 _ {\rm Higgs}/ v _ h ^ 2 $ with $ v _ h = 174 $GeV ($ = v $ in the notation of 
Ref.~\cite{salvio}) which is 4 times larger
than $ \lambda  $. In Ref.~\cite{salvio} the rate depends on $ \lambda  _ h $ the same
way as it does depend on $ \lambda  $ in Ref.~\cite{laine-nlo}.}\label{lambda} 
and $ G _ {\rm F } $ is the Fermi constant.
The neutrino Yukawa couplings are defined in the $ \overline{\mbox{MS}} $ 
scheme with renormalization  scale $  M _ N $.

The leading temperature-dependent term in $ a $ 
is of order 
$ \lambda  T ^ 2/M _ N ^ 2 \sim \lambda  v ^ 4 $, where $ v $ 
is a typical thermal velocity of the heavy neutrinos (cf.\ Sect.~\ref{sec:relativistic}).
If we think of our approach as an expansion in $ v $, and if we
assume $ \lambda   $ to be small, then this term is
of higher order than the one we included in Sect.~\ref{sec:relativistic}. 
The other temperature dependent radiative corrections are of even 
higher
order ($  \sim g ^ 2 v ^ 8 $), where  $ g $ denotes a generic 
Standard Model gauge or Yukawa coupling. It would be straightforward to extend the 
equations in Sect.~\ref{sec:relativistic} to higher orders in $ v $ 
as well. We did not do so, because, as we show in  Sect.~\ref{sec:numerical},
the order $  v ^ 2 $ corrections are quite small in the
parameter region of interest.


\section{Numerical results}
\label{sec:numerical}

In this section we solve the equations for leptogenesis in the 
non-relativistic limit
(Sect.~\ref{sec:non-relativistic}), and we add relativistic (Sect.~\ref{sec:relativistic})
and radiative (Sect.~\ref{sec:radiative}) corrections. We parametrize our results 
by the effective light neutrino mass~\cite{plumacher} 
\begin{align} 
   \widetilde{ m } _ 1 \equiv 
   \frac{ \left ( m _ {\rm D } ^\dagger m _ {\rm D } \right )  _ { 11 }  }
   { M _ N } 
   ,
\end{align} 
where $ m _ {\rm D } $ is the Dirac mass matrix of the neutrinos. 
It is related to the  washout pa\-ra\-meter $ K $ by \cite{buchmuller-pedestrians}
\begin{align}
   \widetilde{m } _ 1  = K m _ \ast
   \label{m1tilde} 
   .
\end{align} 
We have used the Standard Model value
\begin{align}
    m _ \ast \simeq 1.08  \times  10 ^{ -3 } {\rm eV } 
    .
\end{align} 
It  
can be shown \cite{fujii} that $  \widetilde{ m } _ 1 $ is 
larger than the smallest light neutrino mass. 
The range $( \Delta m ^ 2_ {\rm solar } ) ^{ 1/2 } < \widetilde{ m } _
1 < ( \Delta m ^ 2 _ {\rm atmospheric } ) ^{ 1/2 } $ corresponds to $
7.4 < K < 46 $. 

\begin{figure}[t]
  \centerline{
    \epsfxsize=10cm
    \epsffile{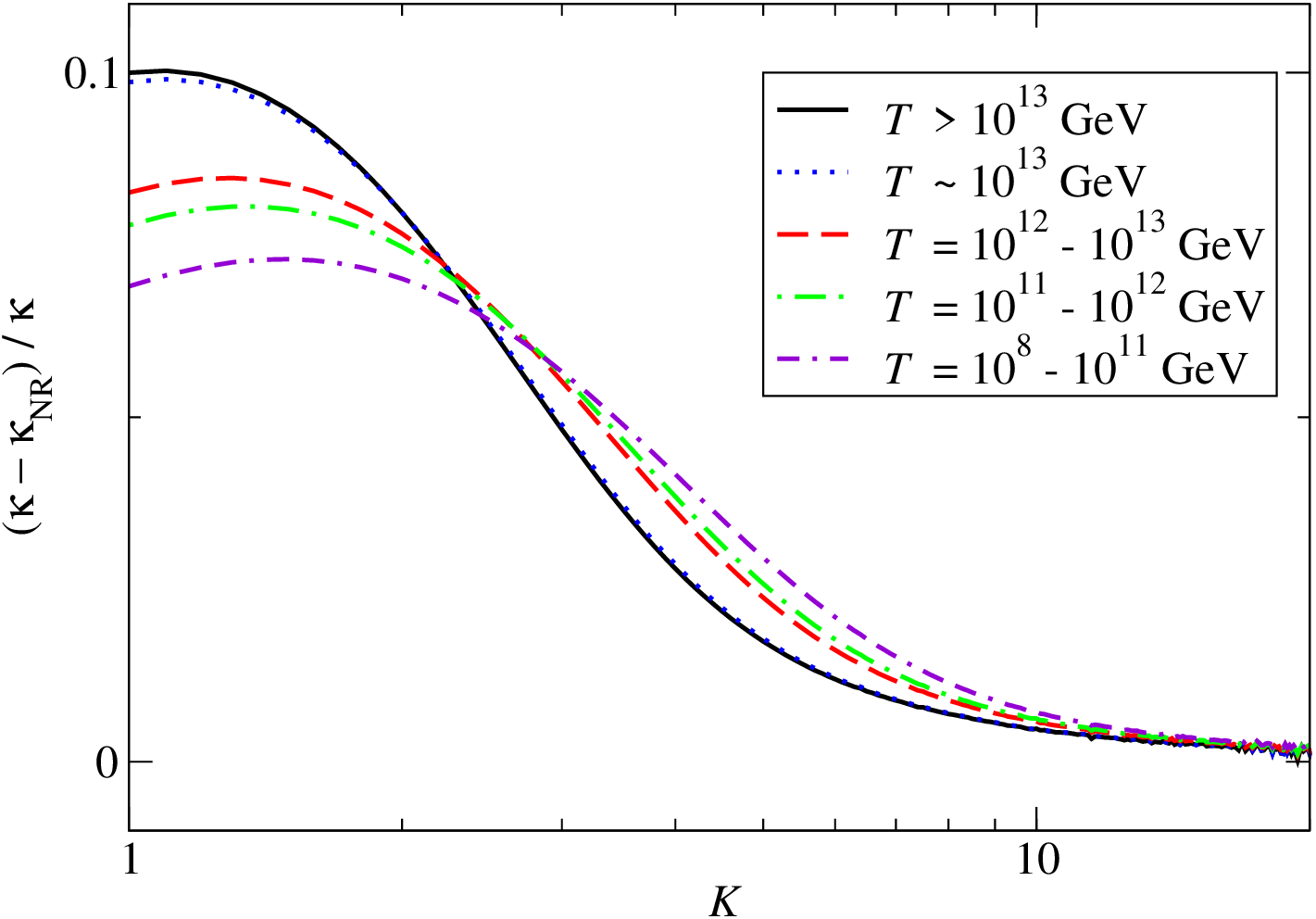}
  } \vspace{0cm}
  \caption[a]{
    Relative size of the relativistic corrections to the 
    $ B - L $ asymmetry. 
Here we used thermal initial conditions. The temperature ranges
correspond to the different values of $ c _ \ell $ and $ c _ \varphi  $. For
$ T \lsim  10 ^{ 12 } $GeV these values correspond to the right-handed neutrinos
 decaying into
$ \tau  $-leptons. 
    \label{fg:K-dependence}
          }
\end{figure}

We have started the evolution of $ n _ N $, $ n _ { B - L } $,  
and $ u $  at $ z = 1 $ with vanishing $ B - L $ asymmetry. Since 
$ v \sim 1 $ for $ z \sim 1 $, the non-relativistic expansion cannot
be expected to work for smaller $ z $. 
Following common practice
we express the final value of $ B - L $ asymmetry in terms of the final 
efficiency factor $ \kappa $ defined by
\begin{align} 
    \lim _ { z \to \infty } 
   \frac{ n _ {B-L} 
   } { n _ \gamma ^ {\rm eq} } 
   \equiv 
   \frac{3}{4} \epsilon \kappa
   \label{kappa} 
   . 
\end{align}
Here $ n _ \gamma ^ {\rm eq} = 2 \zeta  ( 3 ) T ^ 3/\pi  ^ 2 $ 
is the photon equilibrium density, and
$ \epsilon $ is defined in Eq.~(\ref{epsilon}). 

In Fig.~\ref{fg:K-dependence} we show the size of the $ O ( v ^ 2 ) $ 
relativistic 
corrections normalized to the $ O ( v ^ 2 ) $ corrected result 
for several temperature ranges by varying $ c _ \ell $
and $ c _ \varphi $ according to Table 1 of Ref.~\cite{davidson-review}.
We find that for $ K \gtrsim  5 $ the size of the relativistic corrections is
already smaller than $ 3 \% $. A detailed overview of
the results for $ K =  8 $ can be found in Table~\ref{tab:spectators}. 

\begin{figure}[t]
  \centerline{
    \epsfxsize=10cm
    \epsffile{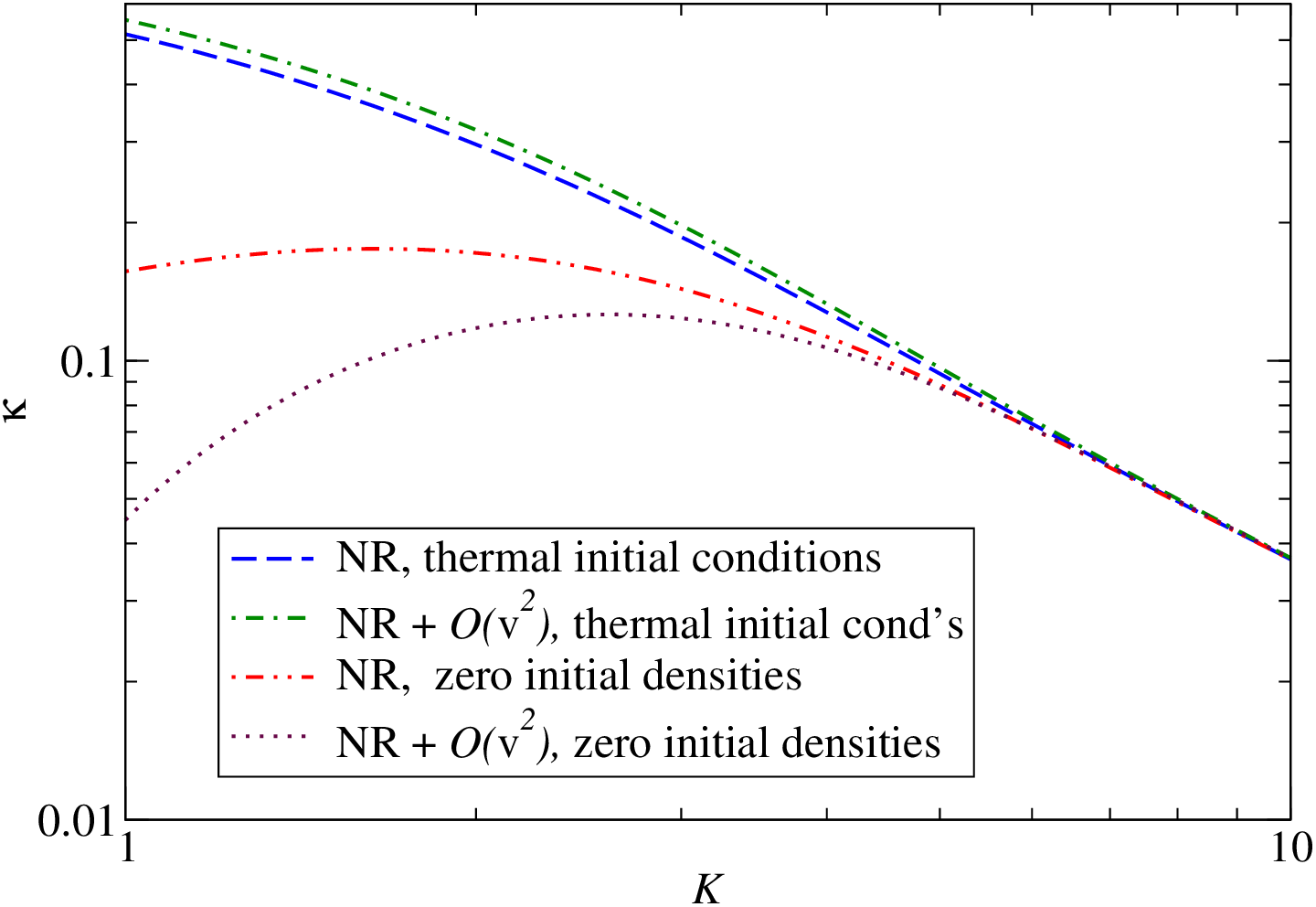}
  } \vspace{0cm}
  \caption[a]{
 Final efficiency factors for two different initial conditions imposed at $ z = 1 $. 
 For thermal initial conditions the non-relativistic approximation remains fairly
 reliable down to very small $ K $. For vanishing initial densities it breaks down
 for $ K \lsim 4 $. For larger $ K $ the dependence on the initial conditions is 
 rather weak. The
 temperature is in the range $ 10 ^ 8 \lsim  T/\mbox{GeV} \lsim  10 ^{ 11 } $. 
    \label{fg:zero}
          }
\end{figure}

Next we study the size 
of the relativistic corrections for different initial conditions for the
temperature range $ 10 ^ 8 \lsim  T/\mbox{GeV} \lsim  10 ^{ 11 } $. We compare 
thermal initial conditions with the extreme case of 
zero initial values for $ n _ N $ and $ u $ 
at $ z=1 $. The latter initial condition is somewhat unphysical, because 
for $ K \gsim  1 $ a substantial number of right handed neutrinos 
will have been thermally produced by the time when $ z=1 $. In Fig.~\ref{fg:zero} 
we compare the results for these two cases. For $ K \lsim 4 $ the non-relativistic
approximation clearly breaks down for zero initial densities. For 
$ K \gsim  5 $ the relativistic corrections
remain small ($ \lsim 1.6\% $ for the parameters in Fig.~\ref{fg:zero}) 
for zero initial values.

It is remarkable that the corrections are so small. This clearly indicates
that the non-relativistic approximation works very well, and that it is not necessary
to use a Boltzmann equation to take into account the momentum distribution of 
the heavy neutrinos. 

\begin{table}
\centering
\renewcommand{\arraystretch}{2}
\begin{tabular}{ccccccc}
 $ T $ (GeV) & spectators & $c_{\ell}$ & $c_{\varphi}$ & $ \kappa_{\rm{NR}} $ & $ \kappa $ & ($ \kappa - \kappa_{\textnormal{NR}} $) / $ \kappa $ \\
 \hline
  $\gg 10^{13}$ & $h_t$, gauge & 1 & $\nicefrac{2}{3}$ & 0.0253 & 0.0255 & +0.7\% \\
  $\sim 10^{13}$ & + QCD sphalerons & 1 & $\nicefrac{14}{23}$ & 0.0260 & 0.0262 & +0.7\% \\
  $10^{12}$ - $10^{13}$ & + $h_b, h_{\tau}$ & $\nicefrac{3}{4}$ & $\nicefrac{1}{2}$ & 0.0363 & 0.0366 & +0.9\% \\
  $10^{11}$ - $10^{12}$ & + $h_c, h_s, h_{\mu}$ & $\nicefrac{78}{115}$ & $\nicefrac{56}{115}$ & 0.0403 & 0.0406 & +1.0\% \\
  $10^{8}$ - $10^{11}$ & + EW sphalerons & $\nicefrac{344}{537}$ & $\nicefrac{52}{179}$ & 0.0495 & 0.0500 & +1.1\% \\
   - & - & 1 & 0 & 0.0363 & 0.0366 & +0.9\%
\end{tabular}
  \caption{Final efficiency factors in the non-relativistic
approximation ($ \kappa  _ {\rm NR } $) and including $ O ( v ^ 2 ) $  corrections
($ \kappa  $). Here we have used $ K = 8 $ and thermal initial conditions. 
The values of $ c_{\ell} $ 
and $ c_{\varphi} $ are taken from Ref.~\cite{davidson-review}. We also include
the case $ c_{\ell} = 1 $, $ c _ {\varphi} = 0 $ which has been used in many papers.}
  \label{tab:spectators}
\end{table}

In our approach a deviation from kinetic equilibrium 
enters the difference $ u - u ^{\rm eq} $. Thus the fact that the relativistic
corrections are so small also indicates that kinetic equilibrium is a good
approximation. We have checked this by comparing the non-relativistic approximation
with the kinetic equilibrium approximation. The difference between the two is 
indeed quite small, and similar 
in size as the $ O ( v ^ 2 )  $  corrections. What turns out to be 
even smaller is the difference
between our non-relativistic approximation plus $ O ( v ^ 2 ) $ corrections
and the kinetic equilibrium approximation. For example, for $ K = 8 $ and $ T= 10 ^ 
{ 10 } $
GeV the kinetic equilibrium approximation deviates from the non-relativistic one
by 0.87\%, and it deviates from the $ O ( v ^ 2 ) $ corrected one by only 0.3\%.

\begin{figure}[t]
  \centerline{
    \epsfxsize=10cm
    \epsffile{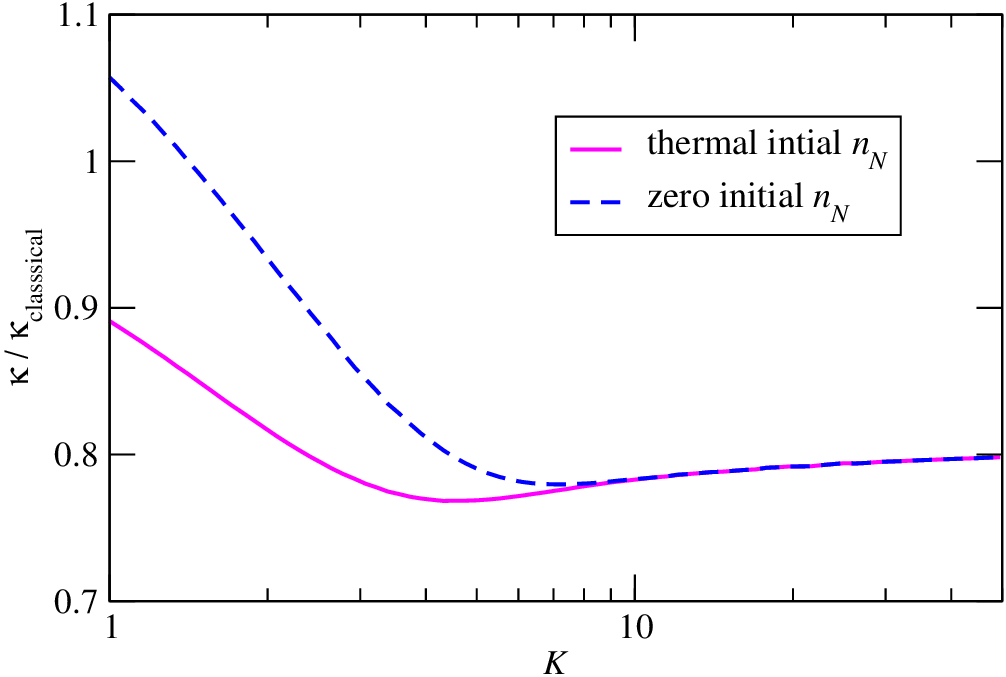}
  } \vspace{0cm}    
  \caption[a]{
    Final efficiency factor $ \kappa  $ 
    computed with Bose- and Fermi-distributions in the washout term, divided
    by the final efficiency factor with the classical 
    Maxwell-Boltzmann statistics. In both cases we 
    have used the non-relativistic
    approximation. We have chosen 
    $ c _ \ell = 1 $, $ c _ \varphi  =0 $ to facilitate comparison with  
    Ref.~\cite{basboll}. This figure shows that it is important to 
    use the correct quantum statistics. Even at very large $ K $ the error
    caused by using classical statistics is of order 20\%. 
    \label{fg:boltzmann}
          }
\end{figure}

It is interesting to compare our results with those 
of Refs.~\cite{basboll,garayoa,hahn}. There Boltzmann equations were 
used to compute the momentum spectrum of the heavy neutrinos without assuming
kinetic equilibrium. 
Ref.~\cite{basboll} compares results obtained with  quantum 
statistics and the momentum dependent Boltzmann equation  with results obtained
with classical statistics and kinetic equilibrium.  
For $ K > 5 $ a discrepancy of about $ 15\% $ was found. Ref.~\cite{garayoa}
considers the same setup and confirms a discrepancy of 20\% for  $ K > 1 $.
Our results, on the other hand, indicate that the deviation from kinetic 
equilibrium should have a much smaller effect, 
which would mean that the discrepancy found in Ref.~\cite{basboll,garayoa}
is due to the statistics. We have checked this switching from quantum to 
Boltzmann-statistics for Eqs.~(\ref{muell}), (\ref{muphi}) in the non-relativistic
approximation. Results are shown in Fig.~\ref{fg:boltzmann}, where we see that
using Boltzmann distributions 
would underestimate  the asymmetry by at least 20\% in the phenomenologically 
interesting region, confirming that the discrepancy is due to the statistics.
Let us stress again that for Eqs.~(\ref{muell}) and
(\ref{muphi}) the correct quantum statistics has to be used even in the non-relativistic
regime,  because  the corresponding momentum integrals are for relativistic particles
and they are saturated at momenta of order $ T $.  

\begin{figure}[t]           
  \centerline{
    \epsfxsize=14cm
    \epsffile{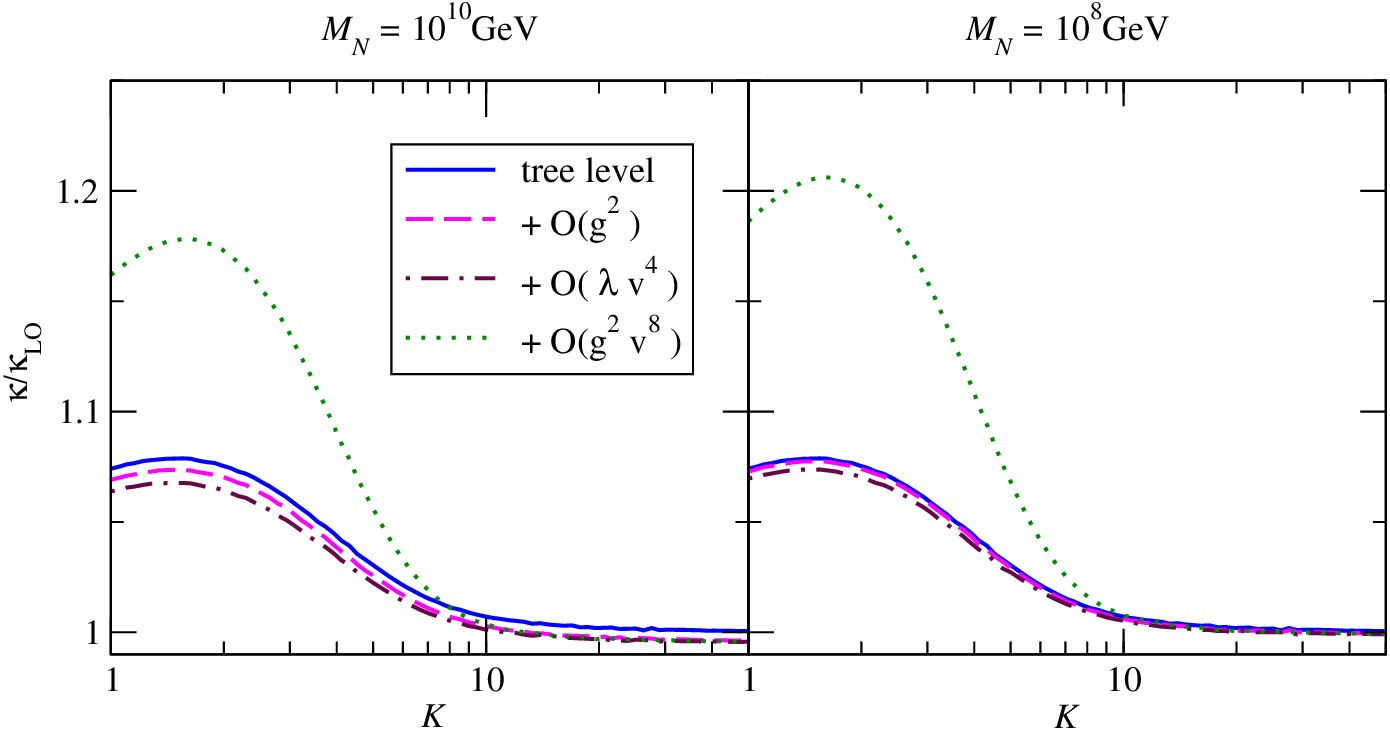}
  } \vspace{0cm}
  \caption[a]{
    Final efficiency factors containing the  
    relativistic $ O ( v ^ 2 ) $ corrections, 
    normalized to the non-relativistic approximation without radiative corrections
    $ \kappa  _ {\rm LO } $. The different curves correspond to 
    different orders included in the coefficients $ \Gamma  _ N $, $ \Gamma  _ { N, u }
    $, and $ \Gamma  _ u  $.
    The renormalization scale of the Standard Model coupling is chosen as the mass 
    of the heavy neutrinos. Thermal initial
    conditions for $ n _ N $ and $ u $ were used. 
    \label{fg:radcorr}
          }
\end{figure}

The radiative corrections 
discussed in Sect.~\ref{sec:radiative}
depend not only on the washout factor, but, through the running
of the Standard Model couplings, also on the neutrino mass $ M _ N $.
In Fig.~\ref{fg:radcorr} we show results for the corrected efficiency factors,
normalized  to the non-relativistic approximation without radiative corrections 
$ \kappa  _ {\rm LO} $, 
for   $ M _ N = 10 ^ { 10 } $  
and $ 10 ^{ 8 } $ GeV. 
All curves contain the relativistic $ O ( v ^ 2 ) $ corrections.
They were obtained with  $ \Gamma _ N $, $ \Gamma _ { 
N, u }  $, and $ \Gamma  _ u $ computed at different orders in the Standard Model
couplings and in $ v ^ 2 \sim T / M _ N $. In the favored 
regime $ 7.4 \lesssim K \lesssim 46 $ the effects of radiative corrections
are 
smaller than the $ O ( v ^ 2 ) $ corrections. For smaller
values of $ K $ the effect of the $ O ( g ^ 2) $ and  the $ O ( \lambda  v ^ 4 ) $ 
corrections to the $ \Gamma  $'s
remain small. 
The  fact that the $ O ( \lambda  v ^ 4 ) $
has a  small effect is partly due to the smallness of the Higgs self-coupling
at these scales. The effect of the
$ O ( g ^ 2 v ^ 8 ) $  contributions grows significantly with decreasing $ K $ (see also 
Table \ref{tab:radcorr}). This indicates
that in the regime where $ v $-dependent corrections
are important, 
the expansion of radiative corrections in powers of $ v ^ 2 $ does not converge 
(cf.\ Ref.~\cite{laine-nlo-relativistic}). 

\section{Summary and  discussion}
\label{sec:discussion}

We have obtained the rate equations for single flavor leptogenesis in the non-relativistic
limit which are valid to all orders in the Standard Model couplings.
We proposed a systematic expansion around the non-relativistic limit.
It can be used to improve the non-relativistic approximation, and to
assess its range of validity. The coefficients in the rate equations  
and the first relativistic corrections were determined
at leading order in the Standard Model couplings. We computed the
$ B - L $ asymmetry from these equations, and we 
found that the relativistic  corrections
are quite small in the parameter region favored by 
atmospheric and solar neutrino oscillations, which lies in 
the so-called strong washout regime. 
They are much smaller than differences
between the non-relativistic approximation and the standard result which 
uses the assumption of kinetic equilibrium for the 
heavy neutrinos as well as classical statistics
for all particle species. 
We found that this difference is mainly due to the statistics used
for computing the washout term, and not so much due
to the assumption of kinetic equilibrium. 

The smallness of relativistic corrections should be very useful, because
one can obtain precise results without having to compute the momentum
spectrum of the right-handed neutrinos, and also 
because it significantly  simplifies the task to take into account Standard Model radiative
corrections. We have taken a step in this
direction by including the production rate of right handed neutrinos
at next-to-leading order in the Standard Model couplings. We find that the radiative
corrections are smaller than the tree-level 
relativistic corrections in the favored region
of the washout factor $ K $. For smaller values of $ K $ they become larger
than the tree-level relativistic corrections. This is due to temperature dependent
gauge and top Yukawa interactions which we consistently included in
a leptogenesis calculation for the first time.

\begin{table}
\centering
\begin{tabular}{c|cc}
  & \multicolumn{2}{c}{$ \left( \kappa_{\textnormal{NLO}} - 
      \kappa _{\textnormal{LO}} \right) /  \kappa_{\textnormal{LO}} $} \\
 \hline
 $ K $ & \small{$ M_N=10^{8} $ GeV} & $ M_N=10^{10} $ GeV \\
 \hline
  5 & $+3.7\%$ & $+2.5\%$ \\
  8 & $+0.5\%$ & $0.0\%$ \\
 10 & $+0.1\%$ & $-0.4\%$ \\
 20 & $-0.1\%$ & $-0.5\%$ \\
 50 & $-0.1\%$ & $-0.5\%$ 
\end{tabular}
 \caption{Influence of the radiative corrections on the final efficiency factor
 for two different values of $ M _ N $. $ \kappa  _ {\rm NLO } $ contains
 all relativistic corrections discussed in Sects.~\ref{sec:relativistic} and
 \ref{sec:radiative} and corresponds to the dotted lines in Fig.~\ref{fg:radcorr}.  
 Thermal initial
   conditions for $ n _ N $ and $ u $ were used. }
 \label{tab:radcorr}
\end{table}

To conclude, the non-relativistic expansion proposed in this paper is a convenient
tool for computing the lepton asymmetry, and to assess the validity of the non-relativistic 
approximation. In case the relativistic corrections are important, also radiative
corrections become important, and then the non-relativistic expansion breaks down. 
In this case one has to use equations which are valid in the fully relativistic regime. 
This conclusion is based on the radiative corrections to the equilibration rate
of right-handed neutrinos. It would be interesting to see whether they are also
valid if radiative corrections to the other terms in the kinetic equations are 
included.

{\bf  Note added:} After this paper was submitted for publication, we were informed by
the authors of Ref.~\cite{biondini}  that they have confirmed the results of 
Ref.~\cite{laine-nlo} for the radiative corrections to the production rate of
right-handed neutrinos.

\section*{Acknowledgements}
 We are grateful to Wilfried Buchm\"uller, Sacha Davidson, and Mikko Laine for 
 helpful discussions, suggestions and comments, and to Nora Brambilla and 
 Antonio Vairo for correspondence.
 This work was supported in part through the DFG funded Graduate School GRK 881.



\begin{thebibliography}{99}

\bibitem{fukugita}  
  M.~Fukugita and T.~Yanagida,
  \emph{Baryogenesis without grand unification,}
  Phys.\ Lett.\  B {\bf 174} (1986) 45. 

\bibitem{buchmuller-origin}
  W.~Buchm\"uller, R.~D.~Peccei and T.~Yanagida,
  {\em  Leptogenesis as the origin of matter,} 
  Ann.\ Rev.\ Nucl.\ Part.\ Sci.\  {\bf 55} (2005) 311
  [hep-ph/0502169].

\bibitem{davidson-review}
 S.~Davidson, E.~Nardi and Y.~Nir,
  {\em  Leptogenesis,} 
  Phys.\ Rept.\  {\bf 466} (2008) 105
  [arXiv:0802.2962 [hep-ph]].

\bibitem{luty}
M.~A.~Luty,
{\em  Baryogenesis via leptogenesis,} 
  Phys.\ Rev.\  D {\bf 45} (1992) 455.

\bibitem{basboll}
A.~Basb\o ll and S.~Hannestad,
  {\em  Decay of heavy Majorana neutrinos using the full Boltzmann equation 
   including its implications for leptogenesis,} 
  JCAP {\bf 0701} (2007) 003
  [hep-ph/0609025].

\bibitem{garayoa}
 J.~Garayoa, S.~Pastor, T.~Pinto, N.~Rius and O.~Vives,
  {\em  On the full Boltzmann equations for Leptogenesis,} 
  JCAP {\bf 0909} (2009) 035
  [arXiv:0905.4834 [hep-ph]].

\bibitem{hahn}
  F.~Hahn-Woernle, M.~Pl\"umacher and Y.~Wong,
  \emph{Full Boltzmann equations for leptogenesis including scattering,}
  JCAP {\bf 0908} (2009) 028
  [arXiv:0907.0205].

\bibitem{salvio} A.~Salvio, P.~Lodone and A.~Strumia,
  {\em  Towards leptogenesis at NLO: the right-handed neutrino interaction rate,}
  JHEP {\bf 1108} (2011) 116
  [arXiv:1106.2814 [hep-ph]].

\bibitem{laine-nlo}
M.~Laine and Y.~Schr\"oder,
  {\em  Thermal right-handed neutrino production rate in the non-relativistic
  regime,} JHEP {\bf 1202} (2012) 068
 [arXiv:1112.1205 [hep-ph]].

\bibitem{laine-nlo-relativistic}
M.~Laine,
{\em  Thermal right-handed neutrino production rate in the relativistic regime,} 
  JHEP {\bf 1308} (2013) 138
  [arXiv:1307.4909 [hep-ph]].

\bibitem{baier}
R.~Baier, Y.~L.~Dokshitzer, S.~Peigne and D.~Schiff,
 {\em  Induced gluon radiation in a QCD medium,} 
  Phys.\ Lett.\ B {\bf 345} (1995) 277
  [hep-ph/9411409].

\bibitem{arnold-stopping}
P.~B.~Arnold, S.~Cantrell and W.~Xiao,
  {\em  Stopping distance for high energy jets in weakly-coupled quark-gluon plasmas,}
  Phys.\ Rev.\ D {\bf 81} (2010) 045017
  [arXiv:0912.3862 [hep-ph]].

\bibitem{beneke-flavored}
 M.~Beneke, B.~Garbrecht, C.~Fidler, M.~Herranen and P.~Schwaller,
{\em  Flavoured Leptogenesis in the CTP Formalism,} 
  Nucl.\ Phys.\ B {\bf 843} (2011) 177
  [arXiv:1007.4783 [hep-ph]].

\bibitem{barbieri}
R.~Barbieri, P.~Creminelli, A.~Strumia and N.~Tetradis,
  {\em  Baryogenesis through leptogenesis,} 
  Nucl.\ Phys.\ B {\bf 575} (2000) 61
  [hep-ph/9911315].

\bibitem{buchmuller-spectator}
 W.~Buchm\"uller and M.~Pl\"umacher,
  {\em  Spectator processes and baryogenesis,} 
  Phys.\ Lett.\ B {\bf 511} (2001) 74
  [hep-ph/0104189].

\bibitem{buchmuller-pedestrians}
W.~B\"uchmuller, P.~Di Bari and M.~Plumacher,
  {\em  Leptogenesis for pedestrians,} 
  Annals Phys.\  {\bf 315} (2005) 305
  [hep-ph/0401240].

\bibitem{plumacher}
M.~Pl\"umacher,
  {\em  Baryogenesis and lepton number violation,} 
  Z.\ Phys.\ C {\bf 74} (1997) 549
  [hep-ph/9604229].

\bibitem{fujii}
M.~Fujii, K.~Hamaguchi and T.~Yanagida,
  {\em  Leptogenesis with almost degenerate majorana neutrinos,} 
  Phys.\ Rev.\ D {\bf 65} (2002) 115012
  [hep-ph/0202210].

\bibitem{biondini}
S.~Biondini, N.~Brambilla, M.~A.~Escobedo and A.~Vairo,
  {\em  An effective field theory for non-relativistic Majorana neutrinos,} 
  JHEP {\bf 1312} (2013) 028
  [arXiv:1307.7680, arXiv:1307.7680].


\end{thebibliography}
\end{document}